\documentclass[12pt,pdftex]{article}
\usepackage{times}

\usepackage[dvips]{graphicx}
\usepackage[usenames,dvipsnames]{color}
\usepackage{amssymb}
\usepackage[pdftex]{hyperref}
\usepackage{amssymb}
\usepackage{wrapfig}
\usepackage{listings}
\usepackage{multirow}

\usepackage[all]{xy}

\usepackage[T1]{fontenc}
\usepackage{calc}
\usepackage{setspace}
\usepackage{color}

\begin{document}

\title{Happiness is assortative in online social networks.}
%\title{@sadpandas: make happy friends on \#Twitter FTW!}
\author{Johan Bollen, Bruno Gon\c calves, Guangchen Ruan, \& Huina Mao}

\maketitle

\begin{abstract}
Social networks tend to disproportionally favor connections between individuals with either similar or dissimilar characteristics. 
This propensity, referred to as assortative mixing or homophily, is expressed as the correlation between attribute values of nearest neighbour vertices in a graph. Recent results indicate that beyond demographic features such as age, sex and race, even psychological states such as ``loneliness''  can be assortative in a social network. In spite of the increasing societal importance of online social networks it is unknown whether assortative mixing of psychological states takes place in situations where social ties are mediated solely by online networking services in the absence of physical contact. Here, we show that general happiness or Subjective Well-Being (SWB) of Twitter users, as measured from a 6 month record of their individual tweets, is indeed assortative across the Twitter social network. To our knowledge this is the first result that shows assortative mixing in online networks at the level of SWB. Our results imply that online social networks may be equally subject to the social mechanisms that cause assortative mixing in real social networks and that such assortative mixing takes place at the level of SWB. Given the increasing prevalence of online social networks, their propensity to connect users with similar levels of SWB may be an important instrument in better understanding how both positive and negative sentiments spread through online social ties. Future research may focus on how event-specific mood states can propagate and influence user behavior in ``real life''.
\end{abstract}

%%%%%%%%%%%%%%%%%%%
\section{Introduction}

As the old adage goes, ``Birds of a feather flock together''. In network theory, this effect is known as \emph{homophily} \cite{McPherson2001} or \emph{assortative mixing} \cite{assort:newman2002,newman03-2,newman:mixpatt2003}, occurs in a network when it disproportionally favors connections between vertices with similar characteristics. The opposite trend, that of favouring connections between nodes with different characteristics, is known as disassortative mixing. For example, a friendship network \cite{McPherson2001} may be highly assortative if it connects individuals who are at similar locations or have similar musical tastes. A heterosexual network \cite{rocha10-1} on the other hand will be highly disassortative since partners will tend to be of the opposite sex.  However, few networks are entirely assortative or disassortative: most will exhibit both properties to some degree depending on the particular characteristic.\\

Social networks can exhibit significant degrees of assortative mixing with respect to a variety of demographic attributes such as sex, race, age, religion and education, including behavioral and health attributes \cite{Ibarra1992,Mollica2003,Christakis2007,Cacioppo2010} and even genotypes \cite{fowler11-1}. Surprisingly, this is also the case for certain psychological states such as loneliness \cite{McPherson2001}. In the latter case individuals preferentially share relations with individuals who report equally elevated levels of loneliness and this homophilic tendency increases over time.\\

Although it is clear that psychological states affect behaviour both online \cite{chmiel10-1} and offline,
the mechanisms through which such states exhibit assortativity and contagion across social bonds are not yet fully understood. However, two different processes are conceiveable: that individuals seek homophilic social relations to share subjective experiences (homophilic attachment), or that the emotional state of an individual can influence that of the people with which he or she interacts (contagion) \cite{Parkinson2009}. While both possibilities are clearly in play in real-world social interactions, it is not clear whether or not they are present in {\it online} social systems which do not necessarily emerge from physical contact or in-person communication \cite{Mason2007a,Huntsinger2009}.\\

The Twitter\footnote{Twitter -- \url{http://www.twitter.com}} microblogging service is a case in point. Twitter users can post brief personal updates of less than $140$ characters at any time. These updates, known as ``tweets'', are distributed to a limited group of ``followers'', i.e.~other Twitter users who have elected to ``follow'' the particular user's tweets \cite{java:2007}. These follower relations are of a fundamentally different nature than their off-line counterparts \cite{kwak:twitter2010}; they are not necessarily reciprocated, i.e. directed, nor modulated and are mostly focused on the exchange of information. In effect, a Twitter Follower relation simply represents the fact that one individual is interested in the content produced by another, without the requirement that the interest be reciprocated. As a simple example, consider the case of celebrities that attract the attention and interest of a large number of people without reciprocating it. This arrangement results in a social network in the form of a directed, unweighted graph which is quite different from naturally occurring social networks in which friendship ties are generally symmetric and vary in strength. As a consequence, one would expect homophily and assortative mixing of emotional states to be absent or fundamentally altered in online social networking environments, in particular those with asymmetric, unweighted connections such as Twitter.\\

However, in spite of the expectation that online environments fundamentally alter social interaction, recent results indicate that personal preferences do indeed exhibit homophilic properties in online environments such as BlogCatalog and Last.fm \cite{Bisgin2010}. Tantalizingly this has also been found the case for  \emph{sentiment}  \cite{Zafarani2010} in LiveJournal\footnote{LiveJournal: \url{http://www.livejournal.com/}}. Given the increasing importance of social networking environments in coordinating social unrest \cite{Mungiu-Pippidi2009} and modulating the public's response to large-scale disasters \cite{Oh2010}, it has become a matter of tremendous interest whether and how online social networking environments exhibit homophily or even contagion on the level of sentiment and mood and how online tools can be leveraged to gain understanding about social behaviour \cite{lazer09-1, lee10-1}.\\

Here we investigate whether and to which degree the general happiness or  Subjective Well-Being (SWB) \cite{subjec:diener2009} of individual Twitter users exhibits assortative mixing.  Several previous works have focused on aggregate \cite{balog06-1, measur:dodd2009,mislove10-1,Bollen2010a,bollen:twitter2011} measurements of mood or emotion in entire communities or systems, but we analyse individual mood state in an online social network. On the basis of a collection of 129 million tweets, we track the SWB levels of 102,009 Twitter users over a 6 months period from the content of their tweets. Each is rated on an emotional scale using a standard sentiment analysis tool. A subsequent assortativity analysis of the Twitter social network then reveals its degree of SWB assortative mixing. Our results indicate that the overall SWB of Twitter users is positive, and highly assortative. In other words, Twitter users are preferentially connected to those with whom they share the same level of general happiness or SWB. Furthermore, tie strength seems to play a significant role in modulating the degree of SWB assortativity.

\section{Data and methods}

We collected a large set of Tweets submitted to Twitter in the period from November $28$, $2008$ to May $2009$. The data set consisted of $129$ million tweets submitted by several million Twitter users. Each Tweet contained a unique identifier, date-time of submission (GMT+0), submission type, and textual content, among other information. Some examples are shown below in Table \ref{tweets_examples}.\\

\begin{table}[h!]
\begin{center}
\begin{tabular}{lp{4.5cm}lp{10.6cm}}
ID		&		date-time				&	type		&	text	\\\hline
1		&		2008-11-28 02:35:48	&	web		&	Getting ready for Black Friday. Sleeping out at Circuit City or Walmart not sure which. So cold out.\\
2		&		2008-11-28 02:35:48	&	web		&	@dane I didn't know I had an uncle named Bob :-P  I am going to be checking out the new Flip sometime soon\\
\multicolumn{3}{c}{$\cdots$}
\end{tabular}
\caption{\label{tweets_examples} Examples of Tweet data collected from  November $28$, $2008$ to May $2009$ for $4,844,430$ users}
\end{center}
\end{table}

We complemented this cross-section sample of twitter activity by retrieving the complete history of over $4$ million users, as well as the identity of all of their followers. The final Twitter Follower network contained  $4,844,430$ users (including followers of our users for which we did not collect timeline information). Armed with the social connections and activity of these users we were able to measure the way in which the emotional content of each users varied in time and how it spread across links.

\subsection{Creating a Twitter ``Friend'' network}

The ``Follower'' network we collected consists of a directed graph $G=(V,E)$ in which $V$ represents the set of all $4,844,430$ Twitter users in our collection, and the set of edges $E \subseteq V^2$  in which each directional edge $v \in E$ consist of the $2$-tuplet $(v_i, v_j)$ that indicates that user $v_i$ follows user $v_j$. By design the Twitter social network is based on ``Follower'' relations which are uni-directional and very easy to establish. As such they form a very minimal representation of possible interaction between those who follow and those who are being followed. In fact, it is quite common for a user $v_i$ to follow a user $v_j$, but for $v_j$ not to follow $v_i$ back. As such, follower relations are not necessarily indicative of any personal relation which may {\it de facto} preclude the establishment of assortative mixing and homophily. We therefore distinguish between mere Twitter ``Followers'' and actual ``Friends'' \cite{Szell2010,huberman08-1} by applying the following transformations to the Twitter follower graph $G$:\\

First, we create a network of Twitter Friend relations from the Follower relations in $G$ by only retaining edges $(v_i,v_j) \in E$ for which we can find a \emph{reciprocal} relation $(v_j,v_i)$, i.e. the set of Friend connections $E^\prime = \{(v_i,v_j): \exists (v_j,v_i) \in E\}$, i.e.~two users only share a Friendship tie if they are both following each other.\\

\begin{figure}[ht!]
	\begin{center}
	\includegraphics[width=8cm]{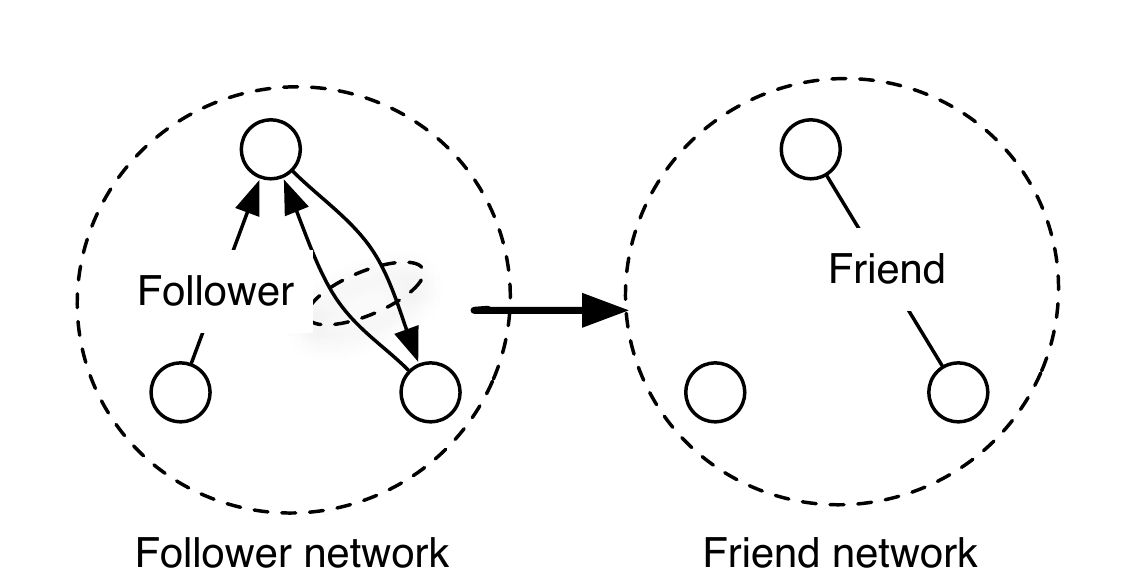}
	\caption{Converting the original Follower network of Twitter into a Friend network by only taking into account reciprocal connections.}
	\end{center}
\end{figure}

Second, to exclude occasional users that are not truly involved in the Twitter social network, we only retained those users in our Twitter Friend network that posted more than $1$ tweet per day on average over the course of $6$ months.\\

Third, we assign a weight $w_{i,j}$ to each edge $(v_i,v_j)$ that serves as an indication of the degree to which users $v_i$ and $v_j$ have similar sets of friends:

\begin{equation}
	w_{i,j} =  \frac{|| C_{i} \cap C_{j}||}{||C_{i} \cup C_{j}||}
	\label{jaccard_friends}
\end{equation}

where $C_{i}$ denotes the neighbourhood of friends surrounding user $v_i$.
Note that this approach does not take into account the number of tweets exchanged between two users, but simply the degree to which two Twitter users have similar friends.\\

\begin{figure}[ht!]
	\begin{center}
	\includegraphics[width=8cm]{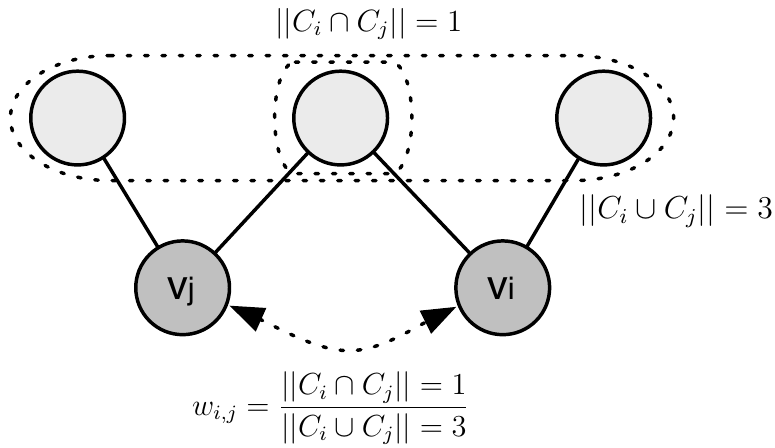}
	\caption{Example of Twitter Friend similarity as calculated according to Eq. \ref{jaccard_friends}. Users $v_i$ and $v_j$ share 1 friend out of three total. Therefore their connection is assigned a weight $w_{i,j}=\frac{1}{3}$.}
	\end{center}
\end{figure}

Finally, we extracted the largest Connected Component ($G_{CC}$) from the resulting network, thereby obtaining a Twitter Friend network of  $102,009$ users and $2,361,547$ edges.\\

\begin{table}[h!]
\begin{center}
\begin{tabular}{l|l}
	Network parameter				&		Values				\\\hline
	Nodes						&		$102,009$ users		\\
	Edges						&		$2,361,547$ edges		\\
	Density						&		0.000454				\\
	Diameter						&		14					\\
	Average Degree				&		46.300				\\
	Average Clustering Coefficient		&		0.262				\\\hline
\end{tabular}
\caption{\label{Gc_parameters} Network parameters for largest Connected Component of Twitter Friend network.}
\end{center}
\end{table}

The reduction in nodes from our original Twitter Follower network ($4,844,430$) to the resulting Friend network ($102,009$) indicates that in Twitter only a small fraction of users are involved in the type of reciprocated Follower type that we considere indicative of actual social relationships. However, once this reduction has occurred, we find that the largest Connected Component of the Friend network, $G_{CC}$, retains $97.9\%$ of users in the original Twitter Friend network. This indicates a high degree of connectivity across all users in the final Friend graph. This is further confirmed by the diameter of $G_{CC}$ which was found to be only $14$ in spite of its low density. Other relevant network parameters for $G_{CC}$ are provided in Table \ref{Gc_parameters}.\\

\begin{figure}[h!]
	\begin{center}
	\includegraphics[width=15cm]{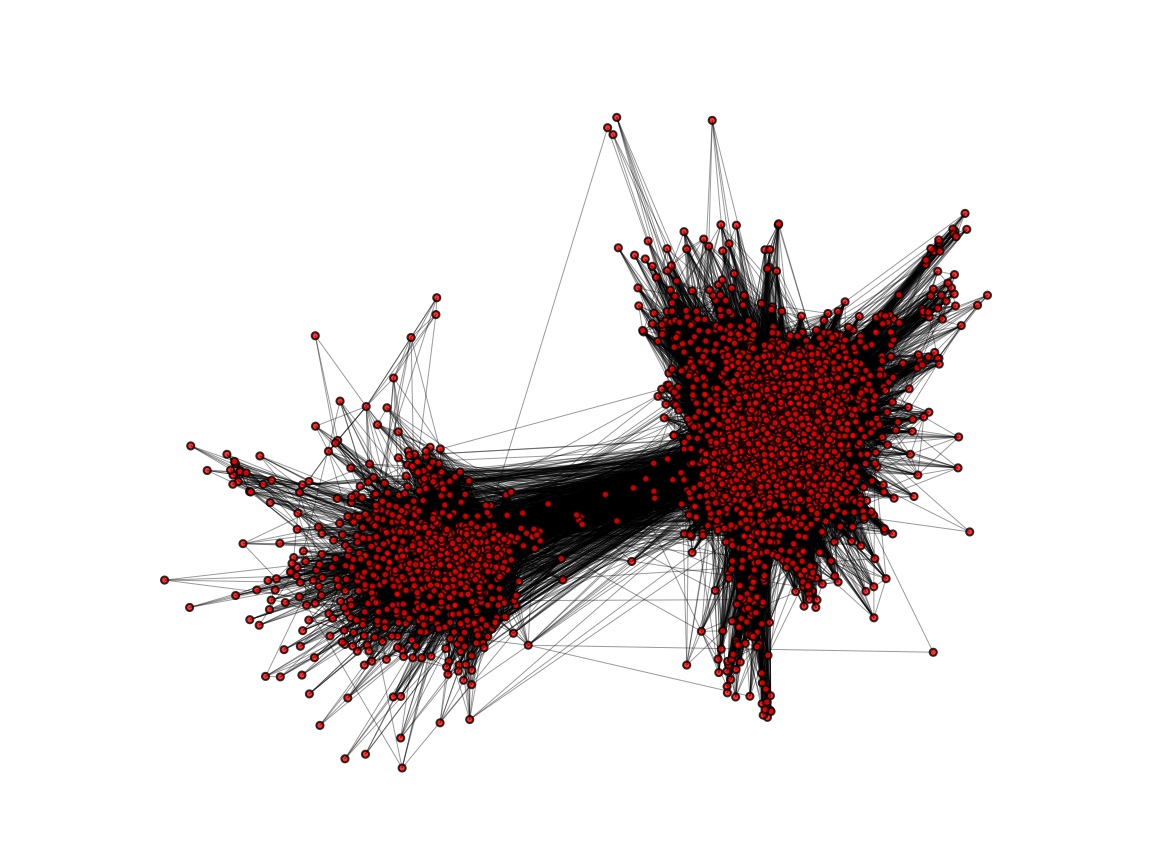}
	\caption{A sub-graph of $3,587$ users extracted from the generated Twitter social network ($102,009$ users and $2,361,547$ edges).}
	\end{center}
\end{figure}

\begin{figure}[h!]
	\begin{center}
	\includegraphics[width=14cm]{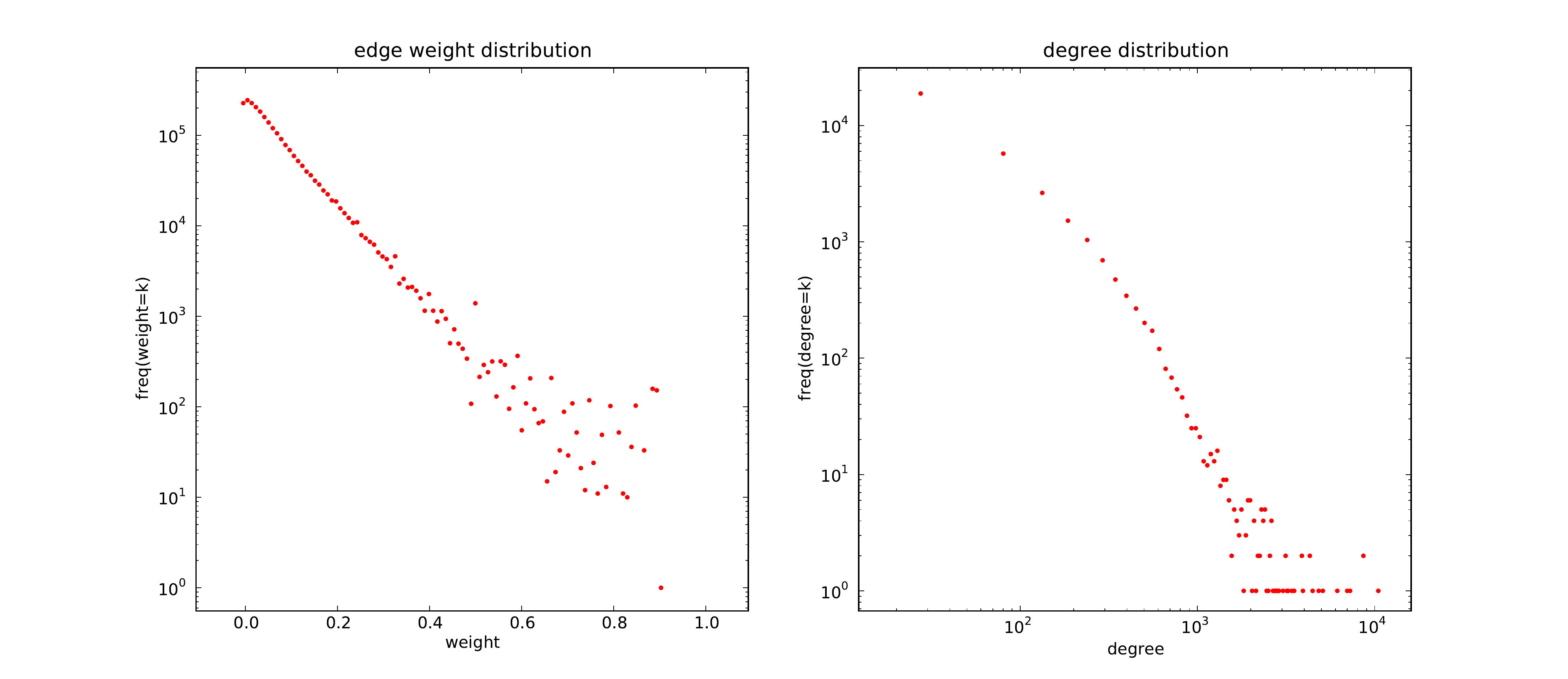}
	\caption{\label{degree_distributions}Twitter Friend network edge weight and degree distributions.}
	\end{center}
\end{figure}

Examining the edge weight distribution as shown in Fig. \ref{degree_distributions} we observe a strongly skewed frequency distribution indicating very many connections in the $G_{CC}$ with low edge weights ($w_{i,j}<0.3$) and few connections with very high edge weights ($w_{i,j}>0.6$). The degree frequency distribution reveals a similar pattern with most users connected to only a few users and a small minority of users connected to thousands of users.

\subsection{User-level measurements of Subjective Well-Being}

We can not directly interrogate Twitter users about their Subjective Well-Being (SWB) \cite{subjec:diener2009}, but we can infer users' SWB from the aggregate emotional content of their tweets over a period of $6$ months. To do so we apply the following procedure.\\

To reduce noise  we only include Twitters users in $G_{CC}$ that posted at least $1$ tweet per day. This guarantees at least $180$ tweets for every individual user from which to assess their SWB. We then analyze the emotional content of each user's 6 month record of tweets using OpinionFinder (OF)\footnote{http://www.cs.pitt.edu/mpqa/opinionfinderrelease/} which is a publicly available software package for sentiment analysis that can be applied to determine sentence-level subjectivity \cite{recogn:wilson2005}. OF has been successfully used to analyze the emotional content of large collections of tweets \cite{tweets:oconnor2010} by using its lexicon to determine the dominance of positive or negative tweets on a given day.  Here we select both positive and negative words that are marked as either ``weak'' and ``strong'' from the OF sentiment lexicon resulting in a list of $2718$ positive and $4912$ negative words. For each tweet in an individual user's 6 month record we count the number of negative and positive terms from the OF lexicon that it contains, and  increase the individual user's score of either negative or positive tweets by $1$ for each occurrence.\\

The Subjective Well-Being ($\mathcal{S}(u)$) of user $u$ is then defined as the fractional difference between the number of tweets that contain positive OF terms and those that contain negative terms:

%\[ S(v_i) = \frac{N_p(v_i) - N_n(v_i)}{N(v_i)} \label{SWBeq}\]
\[ \mathcal{S}(u) = \frac{N_p(u) - N_n(u)}{N_p(u) + N_n(u)} \label{SWBeq}\]

where $N_p(u)$  and $N_n(u)$ represent respectively the number of positive  and negative tweets for user $u$.\\

A number of examples is shown in Table \ref{SWB_examples}.
 
\begin{table}[h!]
\begin{center}
\begin{tabular}{|p{16cm}|}
\hline
\textbf{Tweets submitted by high SWB users ($>0.5$).}\\\hline
So...nothing quite feels like a good shower, shave and haircut...love it\\
My beautiful friend. i love you sweet smile and your amazing soul\\
i am very happy. People in Chicago loved my conference. Love you, my sweet friends\\
@anonymous thanks for your follow I am following you back, great group amazing people\\
\\
\textbf{Tweets submitted by low SWB users ($<0.0$).}\\\hline
She doesn't deserve the tears but i cry them anyway\\
I'm sick and my body decides to attack my face and make me break out!! WTF :( \\
I think my headphones are electrocuting me.\\
My mom almost killed me this morning. I don't know how much longer i can be here.\\\hline
\end{tabular}
\caption{\label{SWB_examples} Examples of Tweets posted by users with very high and very low SWB values.}
\end{center}
\end{table}

\subsection{Defining SWB assortativity}

Having calculated the SWB values of each users, we can now proceed to measure the degree to which the SWB of connected users is correlated.
Intuitively, a person can be emotionally influenced by their friends in two, complementary, ways: influence can come from interacting with specific individuals to which one may attribute more importance \cite{onnela10-1}. We refer to this  first type as ``pairwise node assortativity" since it assesses the degree to which every two pairwise-connected users have similar SWB values. Another possibility is that each individual is influenced by the overall SWB of all of the people it interacts with. We refer to this second type as ``neighborhood assortativity".\\

\begin{figure}[h!]
	\begin{center}
	\includegraphics[width=9cm]{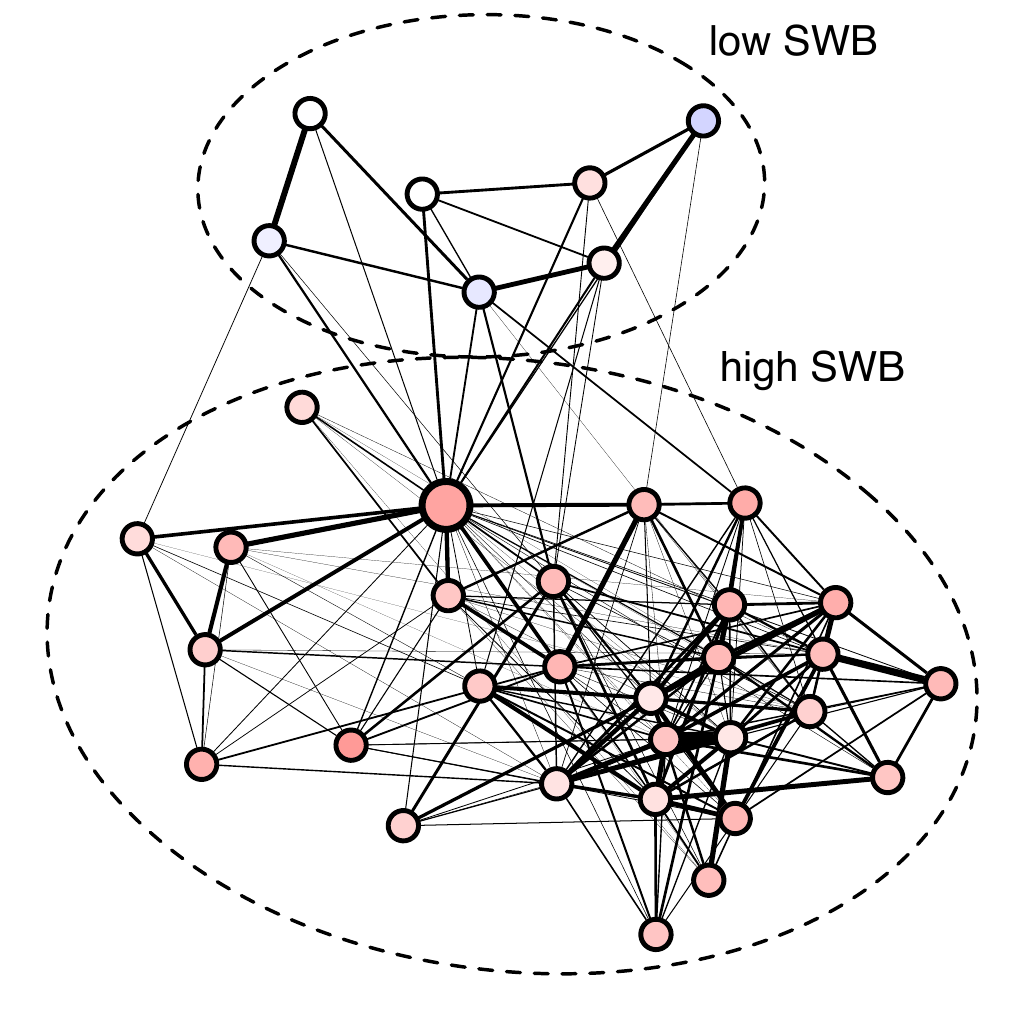}
	\caption{\label{low_SWB_network} Neighborhood network of a very high SWB individual (center). Blue, white, and red node colors correspond respectively to low, neutral and high SWB values. }
	\end{center}
\end{figure}

Fig. \ref{low_SWB_network} illustrates this distinction; it shows the actual neighborhood Friend network of an individual in $G_{CC}$ who has very high SWB values. Nodes are colored according to their SWB values with red indicating high SWB values, blue indicating low SWB values and white indicating neutral or zero SWB values.  The particular individual with high SWB values is connected to a local network of equally high SWB individuals (red). The individual could thus be said to be \emph{neighborhood assortative} within this cluster. However, the individual is also connected to several individuals with low SWB values (blue). For each individual connection this is a case of \emph{pairwise disassortativity}. The cluster of low SWB individuals on the other hand exhibits neighborhood assortativity for low SWB values, and the network in its entirety, including both low and high SWB clusters, exhibits strong SWB assortativity; nodes with similar low or high SWB values tend to be connected (blue and red clusters).\\

We formally define \emph{Pairwise} SWB assortativity as follows: %Each node has been assigned an SWB measure as defined in Eq. \ref{SWBeq}. 
For each edge $(v_i,v_j)$ in the $G_{CC}$ of our social network, we extract the corresponding two SWB values, one for the source node and one for the target node. These values are then aggregated into two vectors, $\mathcal{S}(S)$ and $\mathcal{S}(T)$  for sources and targets respectively.  %Each entry of $s$ represents the SWB value $S(v)$ of that user. 
The value of the pairwise assortativity, denoted $A_{P}(G_{CC})$, is then given by the Pearson correlation coefficient $\rho$ of these two vectors.%  $s_{v_i}$ and $s_{v_j}$ as follows

\begin{equation}
A_{P}(G_{CC})\equiv \rho(\mathcal{S}(S),\mathcal{S}(T))=\frac{1}{n-1}\sum_{i}\left[\left(\frac{\mathcal{S}(S_{i})-\langle \mathcal{S}(S)\rangle}{\sigma\left(\mathcal{S}(S)\right)}\right)\left(\frac{\mathcal{S}(T_{i})-\langle \mathcal{S}(T)\rangle}{\sigma\left(\mathcal{S}(T)\right)}\right)\right]
\label{pairwise_assortativity}
\end{equation}
The pairwise assortativity is then defined in the $[-1,+1]$ interval, with $-1$ indicating perfect disassortativity, $0$ indicates a lack of any assortativity, and $+1$ meaning perfect assortativity.\\

The \emph{neighborhood} assortativity of $G_{CC}$ with regards to SWB, denoted $A_N(G_{CC})$ can be calculated as follows.\\

 For each user $u \in V$, we define its neighborhood:

\begin{equation} 
	\kappa(u) = \{\forall v: \exists (u,v) \in E\}
	\label{neighborhood_def}
\end{equation}

so that $\kappa(u)$ or $\kappa_u$ represents the set of users that user $u$ is connected to. We then calculate an average SWB value for $\kappa(v)$ which we denote 

\begin{equation}
	\overline{\mathcal{S}(\kappa_u)} = \frac{1} {||\kappa(u)||}\sum_{v \in \kappa(u)} \mathcal{S}(v)
	\label{avg_neighborhood}
\end{equation}

We can now define two vectors, one for the SWB values of  every unique user\ and one for the average SWB values of their neighborhoods, denoted by $\mathcal{S}(U)$ and $\overline{\mathcal{S}(\kappa)}$. The neighborhood assortativity of the network $G_{CC}$ with regards to SWB, denoted $A_{\kappa}$, is then given by the correlation function $\rho$ computed over these two vectors as follows:

\begin{equation}
	A_{K} (G_{CC}) \equiv \rho(\mathcal{S}(U),\overline{\mathcal{S}(\kappa)})=\frac{1}{n-1}\sum_{u}\left[\left(\frac{\mathcal{S}(u)-\langle \mathcal{S}(U)\rangle}{\sigma\left(\mathcal{S}(U)\right)}\right)\left(\frac{\overline{\mathcal{S}(\kappa_{u})}-\langle \overline{\mathcal{S}(\kappa)}\rangle}{\sigma\left(\overline{\mathcal{S}(\kappa)}\right)}\right)\right]
	\label{neighborhood_assortativity}
\end{equation}
with the sum to be taken over every user, $u$. $A_{K}(G_{CC})$ then represents the correlation between the SWB values of user $v_i$ and the mean SWB values of its Friends. Similarly to the pairwise version, $A_{\kappa_i}$ is expressed in the range $[-1,+1]$ where $-1$ indicates perfect neighborhood disassortativity and where $+1$ indicates perfect neighborhood assortativity.

%%%%%%%%%%%%%%%%%%%%%%%%%%%%%
\section{Results and discussion}
%%%%%%%%%%%%%%%%%%%%%%%%%%%%%

\subsection{SWB distribution}

\begin{figure}[ht!]
	\begin{center}
	\includegraphics[width=9cm]{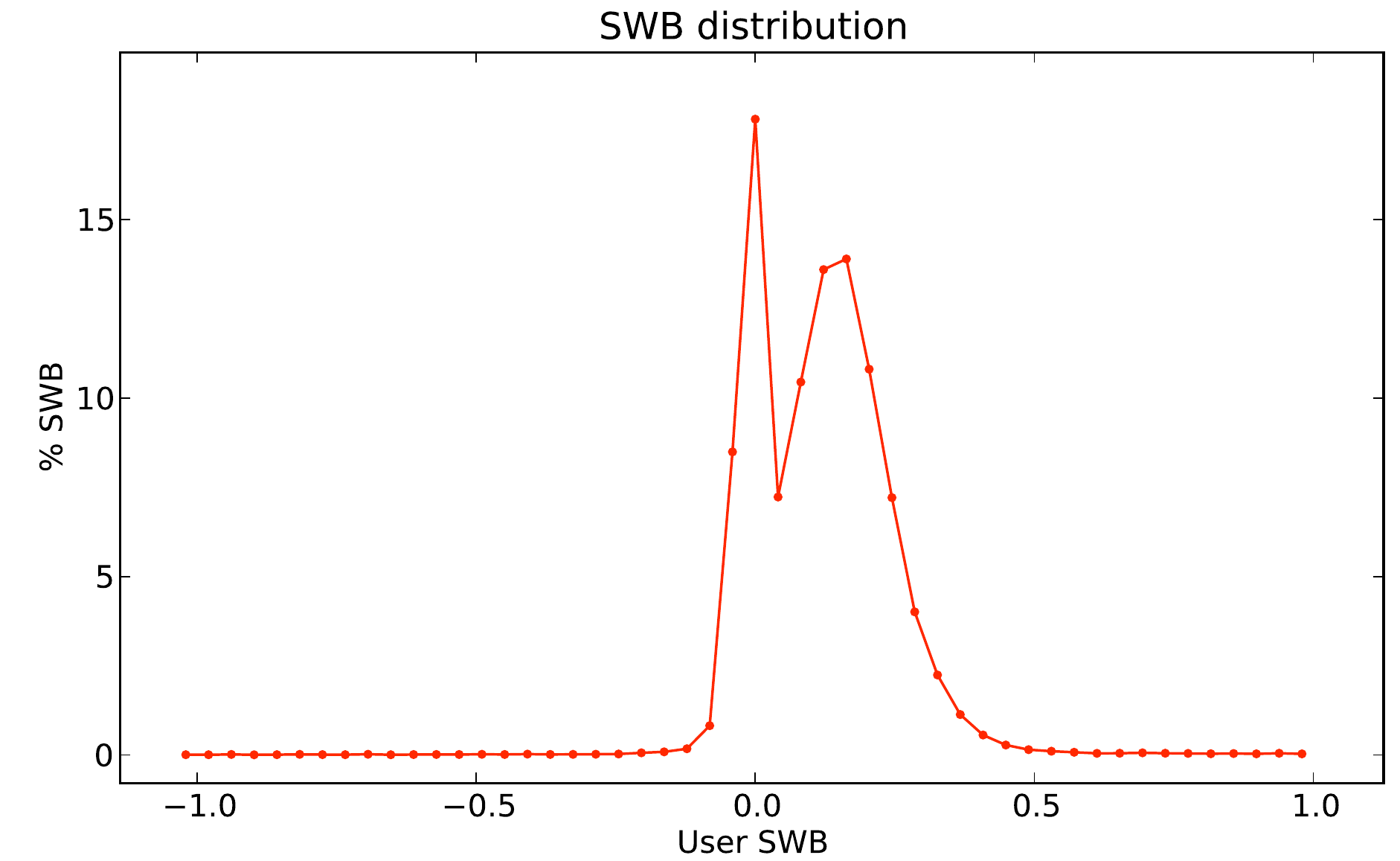}
	\includegraphics[width=9cm]{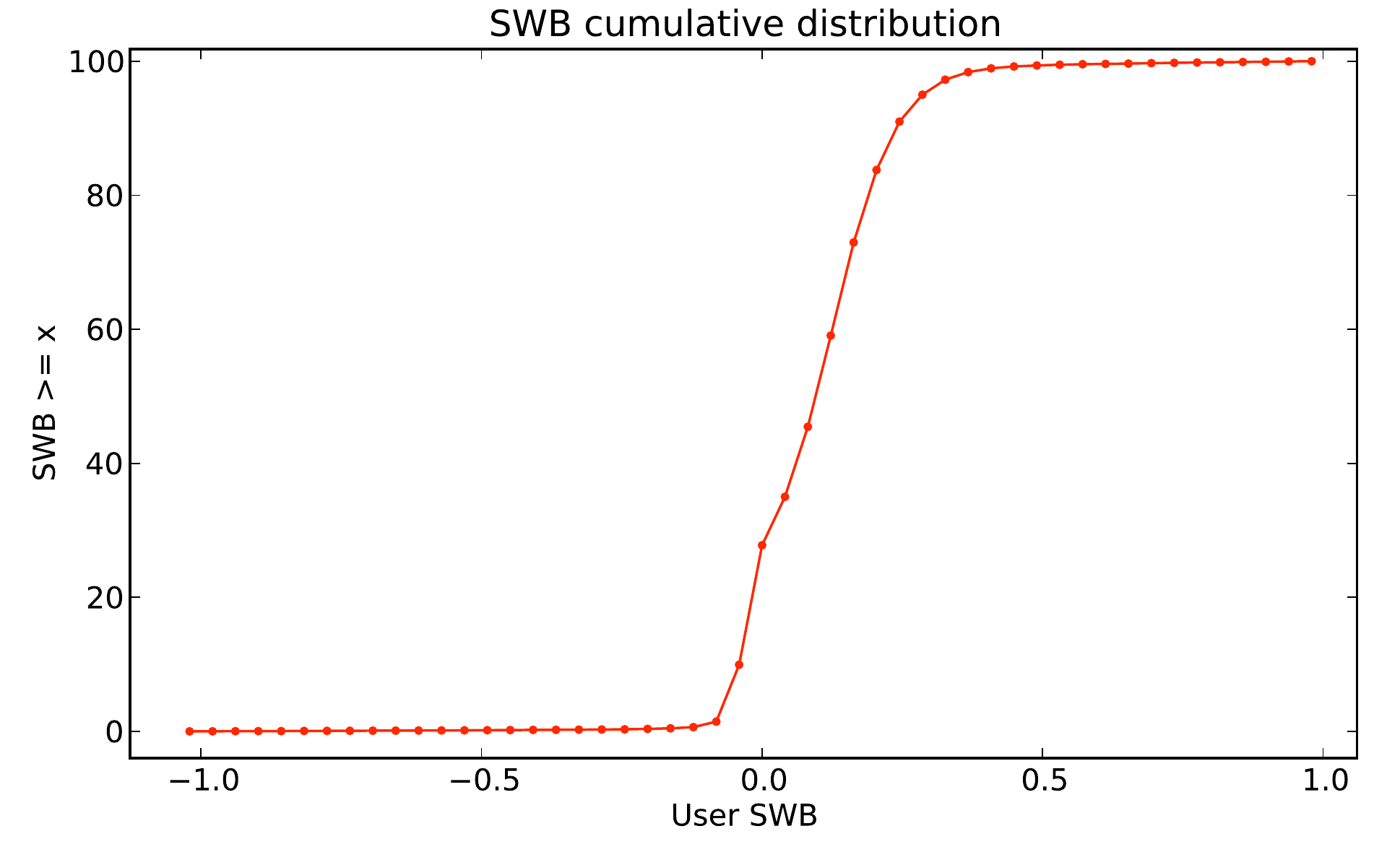}
	\caption{\label{SWB_distributions}Probability distribution (\%) and cumulative distribution (\%) of Subjective Well-Being (SWB) and Emotionality values for our sample of Twitter users.}
	\end{center}
\end{figure}

In Figure~\ref{SWB_distributions} we plot the probability distribution of Subjective Well-Being values across all Twitter users in our sample. The distribution seems bimodal with two peaks: one in the range $[-0.1,0.1]$ and another in the range $[0.2,0.4]$. Excluding users whose SWB=0 (due to a lack of emotional content in their Tweets) we find that a majority of Twitter users in our sample have positive SWB values in a rather narrow range $[0.1, 0.4]$ with a peak at SWB=$0.16$. This is confirmed by the cumulative distribution shown on the left-bottom of Fig. \ref{SWB_distributions}; 50\% of users have SWB values $\leq 0.1$, and 95\% of users have SWB values $\leq0.285$.\\

\subsection{Pairwise and neighborhood SWB assortativity}

In Eq. \ref{pairwise_assortativity} we defined pairwise SWB  as the correlation between the SBW values of connected users in our Twitter Friends networks, whereas Eq. \ref{neighborhood_assortativity} defined neighborhood assortativity was defined as the correlation between the SWB values of individual users and the mean SWB values of their neighbors in the graph $G_{CC}$.\\

The assortativity values were found to be 0.443$^{\star\star\star}$ (N=2,062,714 edges) for the pairwise SBW assortativity and 0.689$^{\star\star\star}$ (N=102,009)\footnote{The sample sizes for pairwise assortativity and neighborhood assortativity) are expressed in edges and nodes respectively, since the former correlation is calculated on the basis of a sample of edges that connect pairs of nodes whereas the other is calculated on the basis of a sample of nodes and their neighborhood} for the neighborhood assortativity. Both correlations are highly statistically significant (p-values $<0.001$) for the sample sizes.\\

\begin{figure}[h!]
	\begin{center}
	\includegraphics[width=18cm]{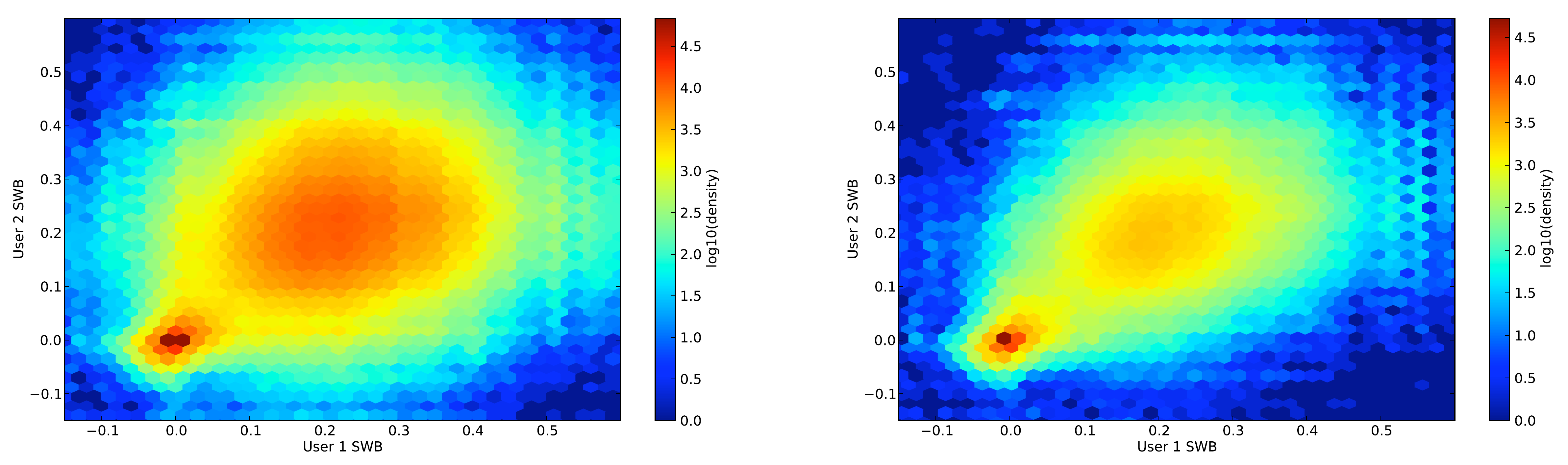}
	\caption{\label{scatter_pairwise}Scatterplot of SWB values for user connected in Twitter Friends network. Left: all edges included. SWB assortativity=0.443$^{\star\star\star}$, N=2,062,714 edges. Right: scatterplot includes edges $w_{i,j} \geq 0.1$, SBW assortativity=0.712$^{\star\star\star}$, N=479,401 }
	\end{center}
\end{figure}

\begin{figure}[h!]
	\begin{center}
	\includegraphics[width=18cm]{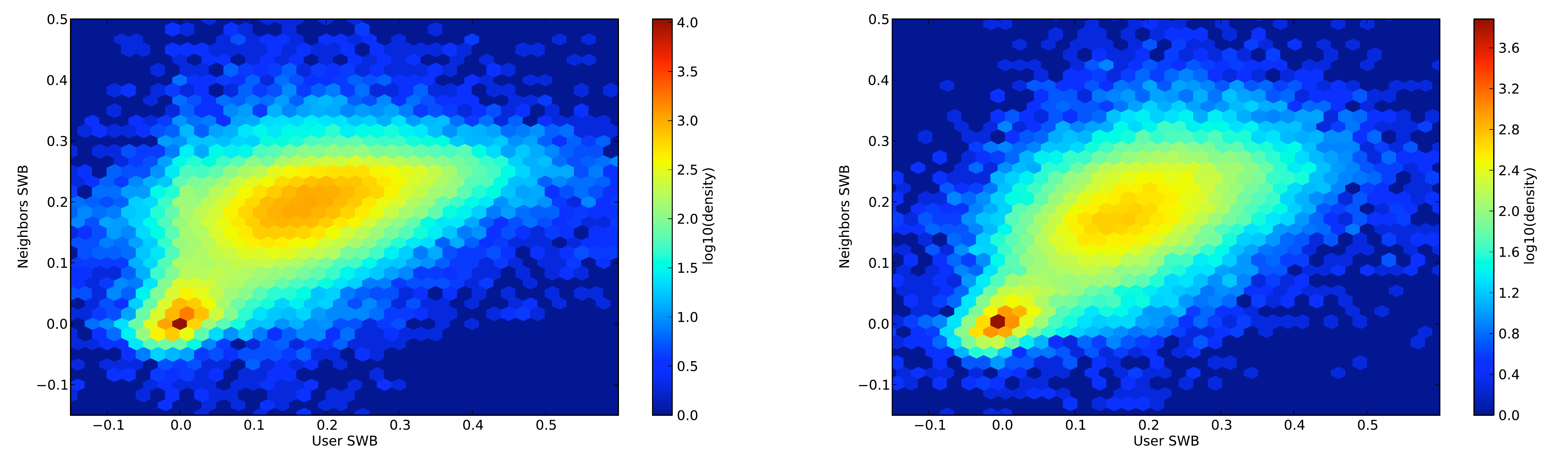}
	\caption{\label{scatter_neighborhood}Scatterplot of SWB values for users (x) and their neighborhood (y) connected in Twitter Friends network.  Left: all edges included. SWB assortativity=0.689$^{\star\star\star}$, N=102,009 nodes. Right: scatterplot includes edges $w_{i,j} \geq 0.1$, SBW assortativity=0.746$^{\star\star\star}$, N=59,952}
	\end{center}
\end{figure}

Regarding pairwise SWB assortativity, the scatterplots on the left of Fig. \ref{scatter_pairwise} and Fig. \ref{scatter_neighborhood} show the distribution of SWB values across the sample of all edges and nodes in $G_{CC}$ and confirm the observed correlation between the SWB values of connected or neighboring users in $G_{CC}$.\\


 The pairwise assortativity scatterplot (Fig. \ref{scatter_pairwise}-left) indicates a significant amount of scatter, commensurate to the lower correlation value of 0.443$^{\star\star\star}$ which is nevertheless statistically highly significant. The observed relation is not obviously linear. The distribution of values is affected by the bimodal distribution of SWB values as shown in Fig. \ref{SWB_distributions}; large numbers of observations cluster at SWB values within the ranges of either $[-0.05,0.05]$ and $[0.1,0.3]$. The clustering pattern of Fig. \ref{scatter_pairwise} however indicates that users with SWB values in a particular range are preferentially connected to users within that same range, thereby confirming the observed positive pairwise SWB assortativity.\\

The neighborhood assortativity scatterplot (Fig. \ref{scatter_neighborhood}-left) indicates a similar effect but here the clustering of users is less pronounced and the amount of scatter is lower than that observed for the pairwise assortativity scatterplot, commensurate to the higher neighborhood assortativity value of 0.689$^{\star\star\star}$. Although less pronounced the bimodal distribution of SWB values is apparent and leads to a clustering of user and neighborhood SWB values in the ranges of $[-0.05,0.05]$ and $[0.1,0.3]$. Nevertheless it is again the case that users with SWB values in either range are most likely to be connected to users or neighborhoods with SWB values in the same range. The distribution of user and neighborhood SWB value is furthermore in line with a positive, linear relationship.

\subsection{Edge weight and SWB assortativity}

Pairwise SWB assortativity and neighborhood SWB assortativity diverge significantly ($0.443^{\star\star\star} < 0.689^{\star\star\star}$) . The former is based on the pairwise comparison of SWB values across all connection in $G_{CC}$ many of which may be weak or irrelevant connections from the perspective of indicating actual Friend ties. To measure the effect of edge weights, we calculate pairwise and neighborhood assortativity values under different edge thresholds, i.e.~we only take into account edges in $G_{CC}$ whose weight as defined in Eq. \ref{jaccard_friends} is $w_{i,j} \geq \epsilon$ where $\epsilon \in [0,1]$ represents a given edge threshold. The consequent assortativity calculations will therefore more strongly reflect only those connections between users that are indicative of stronger Friend relations (higher $w_{i,j}$). In other words we are verifying whether stronger user relations lead to higher or lower assortativity.\\

\begin{figure}[h!]
	\begin{center}
	\includegraphics[width=14cm]{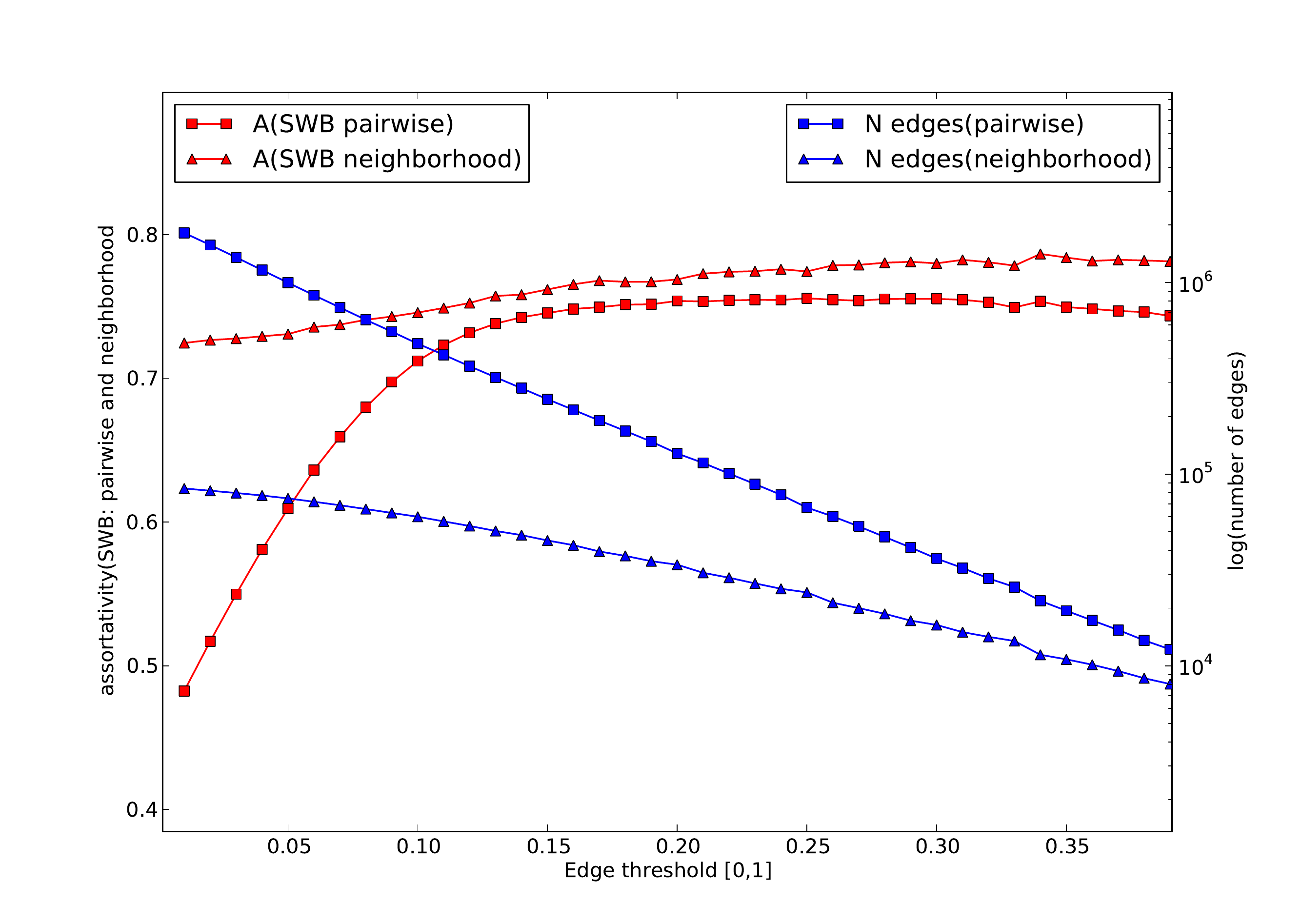}
	\caption{\label{assortativity_thresholds} Pairwise Subjective Well-Being assortativity and log(number of edges) values vs. edge weight thresholds.}
	\end{center}
\end{figure}

The results of the calculation of pairwise and neighborhood SWB assortativity under various edge thresholds are shown in Table \ref{assortativity_edgethreshold} and visualized in Fig. \ref{assortativity_thresholds}. Values for $\epsilon>0.8$ are excluded since the correlation coefficients were not statistically significant (p-value $< 0.1$). The graph in Fig.  \ref{assortativity_thresholds} overlays the different pairwise and neighborhood SWB assortativity values along with the number of remaining edges and nodes under the given edge threshold, i.e.~the sample size for the given assortativity calculation.\\

\begin{table}[h!]
\begin{center}
\begin{tabular}{||r||l|r||l|r||}
			\multicolumn{3}{r}{Pairwise}					&	\multicolumn{2}{r}{Neighborhood}					\\\hline
Edge threshold ($\epsilon$)		&	A(SWB) 					&	N edges		&	A(SWB)					&	N nodes		\\\hline
0.0		 	&	0.443$^{\star\star\star}$		&	2,062,714		&	0.689$^{\star\star\star}$		&	102,009		\\
0.10			&	0.712$^{\star\star\star}$		&	479,401		&	0.746$^{\star\star\star}$		&	59,952		\\
0.20			&	0.754$^{\star\star\star}$		&	128,261		&	0.769$^{\star\star\star}$		&	33,693		\\
0.30			&	0.755$^{\star\star\star}$		&	36,255		&	0.780$^{\star\star\star}$		&	16,334		\\
0.40			&	0.743$^{\star\star\star}$		&	10,355		&	0.779$^{\star\star\star}$		&	7,699		\\
0.50			&	0.757$^{\star\star\star}$		&	3,255		&	0.781$^{\star\star\star}$		&	3,793		\\
0.60			&	0.798$^{\star\star\star}$		&	1,375		&	0.805$^{\star\star\star}$		&	1,439		\\		
0.70			&	0.755$^{\star\star\star}$		&	 689			&	0.816$^{\star\star\star}$		&	502			\\
0.80			&	0.434$^{\star\star\star}$		&	301			&	0.768$^{\star\star\star}$		&	149			\\
0.90			&	-						&	-			&	-						&	-			\\\hline
\end{tabular}
\caption{\label{assortativity_edgethreshold} Pairwise and Neighborhood Subjective Well-Being assortativity values $A(SWB)$ vs. edge threshold $\epsilon$. $^{\star\star\star}$: p-value $<0.001$.}
\end{center}
\end{table}

Both pairwise and neighborhood assortativity values increase as the edge threshold $\epsilon$ increases, but not in a linear manner. Pairwise SWB assortativity values increase sharply as $\epsilon$ increases from $0$ to  $0.10$ and afterwards stabilizes at a value of approximately $0.750$ which is maintained in the interval $\epsilon \in [0.15,0.85]$. In other words, removing edges with $w_{i,j} < 0.1$ increases pairwise SWB assortativity considerably, but the removal of edges with higher $w_{i,j}$ values has little to no additional effect. We observe that $\epsilon= 0.1$ reduces the number of edges by a fifth, namely from 2,062,714 to 479,401 indicating that a large number of edges are characterized by low similarity values which when included lead to lower pairwise SWB assortativity.\\

The neighborhood SWB assortativity increases with higher $\epsilon$ values but in a less pronounced manner. At $\epsilon=0$ we find a neighborhood SWB assortativity value of $0.689$ which increases to approximately $0.760$ for $\epsilon \in [0.10, 0.90]$. We observe significant declines in the number of nodes that remain under increasing $\epsilon$ values as was the case for pairwise assortativity.\\

At a value of $\epsilon=0.1$ we find the highest pairwise and neighborhood SWB assortativity values combined with the largest sample sizes, excluding no threshold i.e.~$\epsilon=0$. We therefore generate new scatterplots of SWB values for pairwise and neighborhood SWB assortativity at $\epsilon=0.1$ as shown in Fig. \ref{scatter_pairwise} (right) and Fig. \ref{scatter_neighborhood} (right). The respective scatterplots reflect higher assortativity values; we find less scatter, a stronger positive and linear relation between SWB value of connected users, and a less pronounced clustering caused by the bimodal distribution of SWB values.\\

\subsection{Discussion}

The above outlined results indicate the following.\\

First, Twitter users in general exhibit a low to moderate SWB, with very few users being characterized by low SWB values. The Twitter population in our sample can therefore in the mean be considered moderately happy. This observation is most likely an underestimation given the relative preponderance of negative terms in the OF lexicon. However, the SWB distribution is bi-modal showing a group of Twitter users with zero to very low SWB values, i.e.~those that are on the average somewhat happy, and another group with more pronounced, higher SWB values. This may result from socio-cultural differences in how emotions and mood are expressed on Twitter. Some users may infrequently express their emotional states whereas other are more prone to do so.\\

Second, we find statistically significant levels of  pairwise and neighborhood SWB assortativity indicating that Twitter users either prefer the company of users with similar SWB values (homophilic attachment) or converge on their Friends' SWB values (contagion). The relation between user SWB values is not linear and biased by the bi-modal distributions of SBW values causing users to be clustered in two groups with equally low or high SWB values. In other words, low SWB users are connected to low SWB and high SWB users are connected to high SWB users. Again, this may confirm the notion that distinct socio-cultural factors affect the expression of emotion and mood on Twitter, and cause users to cluster according to their degree of expressiveness as well as SWB.\\

The results of measuring pairwise and neighborhood assortativity under different edge weights indicate that we find a stronger and more significant relation between the SWB values of connected users when we only take into account connections with higher $w_{i,j}$ weights, i.e.~those that are deemed more reliable indicators of actual Friend ties. A possible mechanism to explain the difference in magnitude between pairwise and neighborhood assortativity might be that users' neighborhoods contain individuals that they are indeed strongly assortative with, and whose SWB values affect mean neighborhood SWB values and thus neighborhood assortativity overall, whereas they are ``drowned out'' in the process of making pairwise comparisons between all individuals that a user is connected to in the process of calculating pairwise SWB assortativity.\\

For example, users might generally have 10 neighbors, but are generally highly SWB assortative with only 1 Friend. In the calculation of pairwise assortativity, this leads to 10 pairwise comparisons between SBW values only one of which contributes to the overall observed pairwise SWB assortativity in the graph. However, the neighborhood assortativity relies on a mean SWB value calculated for the entire neighborhood, including the 1 highly assortative individual. The latter thus influences the average SBW value for the entire neighborhood causing an increased neighborhood assortativity value.\\

The greatest improvement in assortativity values indeed occurs for pairwise SWB assortativity which is most affected by the preponderance of weakly weighted connections since it is defined at the level of all individual user to user connections. Both pairwise and neighborhood SWB assortativity converge on a value of approximately $0.750$ which indicates a significant degree of SWB assortativity in our Twitter Friend graph $G_{CC}$. 

\section{Conclusion}

Recent findings show that assortative mixing can occur in a variety of social contexts and personal attributes. Here we show that Subjective Well-Being is equally assortative in the Twitter social network, i.e.~the SWB of individuals that have reciprocal Twitter follower links are strongly related. Happy users tend to connect to happy users whereas unhappy users tend to be predominantly connected to unhappy users. The convergence of pairwise and neighborhood assortativity under increasing edge weight thresholds indicates that users tend to be most assortative with a limited number of individuals that they have strong social ties to and that weaker ties fulfill a different social role possibly as outlined by Centola et al (2007)\cite{Centola2007}.\\

We do not not address the social or cognitive mechanisms that cause the observed SWB assortativity. Two different mechanisms may be at work\cite{Aral2009}. The first is based on the notion of ``homophily'', i.e.~users and connections tend to preferentially connect to users with similar SWB values. As an online social network grows, new connections are thus biased towards connecting individuals with similar SWB values.  This process may be modeled in terms of ``preferential attachment'' theory. The second mechanism that may cause SWB assortativity is that of ``mood contagion'', namely that connected users converge to similar SWB values over time. In other words, being connected to unhappy users can make one unhappier and vice versa. The latter suggests that users may control their own level of SWB by choosing the right set of online friends and influence their Friends' SWB by creating strong social ties and hoping for some form of SBW contagion to take place. A third possibility is that users asses or express their SWB relative to that of their friends. As a user's neighborhood becomes happier, this may affect their own expression of SWB-related sentiment. This phenomenon may occur at the level of entire cultures which may be comparatively more or less prone to open expressions of individual sentiment.\\

At this point our research does not offer any information on which of these mechanisms cause the observed SWB assortativity or in fact whether both may be occurring. Future research will therefore focus on analyzing user connections and SWB values over time, and relating these changes in the framework of homophily and preferential attachment\cite{homoph:shalizi2011}. Twitter has now become a major international phenomenon, and this investigation must therefore include linguistic, cultural and geographic factors.

\section*{Acknowledgements}
We are grateful to Eliot Smith, Peter Todd and Ishani Banerji for their very useful feedback and input throughout the research that led to this paper. This research was supported by NSF Grant BCS \#1032101.

\bibliographystyle{unsrt}
\bibliography{mood}

\end{document}